\documentclass[11pt]{article}
\usepackage{graphicx}
\usepackage{amsmath}
\usepackage{amssymb}
\usepackage{amsxtra}

\title{Main properties of new heavy hadrons and\\ the luminosity of hadronic dark matter}
\author{V.I. Kuksa\\Research Institute of Physics, Southern Federal University,\\
344090 Rostov-on-Don, Av. Stachky 194, Russian Federation, vkuksa47@mail.ru\\V.A. Beylin\\
Research Institute of Physics, Southern Federal University,\\ 344090 Rostov-on-Don, Av. Stachky 194, Russian Federation}

\begin{document}
\maketitle

\begin{abstract}
The origin and main properties of new heavy hadrons as dark matter candidates, are represented. Low-energy interactions of new hadrons with leptons and nucleons are described in the
terms of effective vertexes. We consider the lowest excited levels of new mesons in the frame-work of the Heavy Quark Effective Theory. The effect of fine and hyper-fine splitting of
excited states follows directly from this theory. We analyze phenomenological consequences of this effect as manifestation of dark matter particles.
\end{abstract}


% optionally

\section{Introduction}\label{s:intro}

The cald dark matter candidates usually are interpreted as stable weakly interacting massive particles (WIMP). Rigid experimental constraints on the cross-section of WIMP-nucleon
interaction \cite{Aprile17} exclude some variants of WIMPs. So, alternative scenarios are considered in literature, for example, the scenario with strongly interacting massive particle
(SIMP) \cite{Belotsky05}-\cite{Bazhutov17}. In these works, the scenario of hadronic dark matter realization was represented, where dark matter (DM) particles consists of new heavy and
ordinary quarks. Such scenarios can be realized in the SM extensions with fourth generation \cite{Belotsky05,Cudell15}, in the chiral-symmetric models \cite{Bazhutov17,Beylin19}, and
in the extension with singlet quark \cite{Beylin18}.

  Principal properties of hadronic DM particles of meson type were considered in Refs.~\cite{Beylin19,Beylin18,Beylin19a}, where it was shown that hadronic DM scenario is not excluded
by EW and cosmochemical constraints. Low-energy interaction of hadronic dark matter (HDM) with ordinary matter was decribed in Refs.~\cite{Beylin20,Kuksa20}. There, Lagrangians of
strong and weak interactions of new heavy mesons with ordinary light ones and gauge bosons were derived. It was shown in \cite{Kuksa20}, that the effect of fine and hyperfine splitting
manifests itself in the spectrum of new heavy mesons. Note, the existence of new heavy hadrons and their principal properties are the direct coseuquenses of high energy SM extensions.
These extensions have independent meaning as variants of realization of grand unification theory. Application of this scenario to the description of DM is not obligatory, however, it
gives the simplest and natural realization of hadronic DM scenario.

 In this report, we consider in details the main properties of new heavy mesons and their phenomenological consequences.
In Section 2, we describe the interaction of new quarks with the gauge bosons, electro-weak restrictions on the mass of these quarks and define the value of their mass. Low-energy
interaction of new mesons with ordinary particles (leptons and nucleons) are considered in Section 3. The effects of fine and hyperfine splitting in the set of new heavy mesons are
described in Section 4. Some conclusions are presented in Section 5.

\section{New Heavy Stable Quarks}

 In the scenario with chiral extension of SM, new sets of the up and down quarks has the form:
\begin{equation}\label{2.1}
Q=\{Q_R=\begin{pmatrix}
  U\\
  D,
  \end{pmatrix}_R\,;
  \,\,\,U_L,\,\,\,D_L\}
\end{equation}
 The structure of covariant derivatives is defined in standard way:
\begin{align}\label{2.2}
D_{\mu}Q_R=&(\partial_{\mu}-ig_1 Y_Q V_{\mu}-\frac{ig_2}{2} \tau_a V^a_{\mu}-ig_3 t_iG^i_{\mu}) Q_R;\notag\\
D_{\mu}U_L=&(\partial_{\mu}-ig_1 Y_U V_{\mu}-ig_3 t_iG^i_{\mu}) U_L,\notag\\D_{\mu}D_L=&(\partial_{\mu}-ig_1 Y_D V_{\mu}-ig_3 t_iG^i_{\mu}) D_L.
\end{align}
In the Eqs.~ (\ref{2.2}), the values $Y_A,\,A=Q,U,D$, are the hypercharges and $t_i$ are generators of $SU_C(3)$ -group. Here, gauge boson fields $V^a_{\mu}$ are superheavy chiral
partner of standard fields.  If we interprete the gauge field $V_{\mu}$ as standard $U(1)_Y$ one, then standard mixing of $V_{\mu}$ and $V^3_{\mu}$ is forbidden. Moreover, standard
interpretation of the field $V_{\mu}$ and weak hypercharge $Y_Q=\bar{q}$ leads to wrong $V-A$ structure of photon interaction. These obstakles were considered in detail in
Ref.~\cite{Beylin19}, where hypercharge operator was redefined and vector-like interaction of new quarks was established:
\begin{equation}\label{2.3}
L_Q^{int}=g_1 V_{\mu}\bar{Q}\gamma^{\mu}\hat{q} Q = g_1 (c_w A_{\mu}- s_w Z_{\mu}) (q_U\bar{U} \gamma^{\mu} U - q_D\bar{D} \gamma^{\mu} D),
\end{equation}
where the field $V_{\mu}$ is standard mixture of photon, $A_{\mu}$, and  boson, $Z_{\mu}$. In expression (\ref{2.3}), the values $c_w=\cos\theta_w$, $s_w=\sin\theta_w$, $g_1 c_w=e$ and
$\theta_w$ is Weinberg angle.

In the scenario with singlet quark (SQ), new heavy quark, $S$, is a singlet with respect to $SU_W(2)$ weak group. The high-energy origin and low-energy phenomenology of singlet quark
(SQ) were considered in many works (see, for example, \cite{Barger86}-\cite{Rattazzi90} and references therein). The low-energy phenomenology of SQ mainly is stipulated by effect of
it's mixing with ordinary quarks. This mixing causes the appearence of flavor changing neutral currents (FCNC) and instability of SQ. Here, we consider the variant with stable SQ which
have no the mixing with ordinary ones. Because the SQ strongly interacts with ordinary quarks, they form the bound states of type $(Sq)$, $(Sqq)$, $(SSq)$. Here, we consider the main
properties of two-quark, $(Sq)$, meson states and the scenario, where the lightest neutral meson $M^0=(\bar{S}q)$ is the DM candidate.

Further, we present the scenario of the SM extension with singlet quark $S$, which can be up, $U$, or down, $D$, type. The field $S$ is singlet representation of $SU_W(2)$ group and
has standard properties with respect to $U_Y(1)$ and color $SU_C(3)$ groups. Minimal Lagrangian of SQ interaction with the gauge bosons is:
\begin{equation}\label{2.4}
L_S=i\bar{S}\gamma^{\mu}(\partial_{\mu}-ig_1 q V_{\mu} -ig_s t_a G^a_{\mu})S - M_S \bar{S} S.
\end{equation}
In (\ref{2.4}), hypercharge $Y/2=q$ of singlet quark $S$, $t_a=\lambda_a/2$ are generators of $SU_C(3)$ -group, and $M_S$ is mass of quark. Abelian part of the Lagrangian (\ref{2.4})
describes the interactions of SQ with photon $A$ and $Z$-boson:
\begin{equation}\label{2.5}
L_S^{int}=g_1 q V_{\mu}\bar{S}\gamma^{\mu} S = q g_1 (c_w A_{\mu}- s_w Z_{\mu})\bar{S}\gamma^{\mu} S.
\end{equation}
Note, the interaction of SQ with $Z$-bosons has vector-like form.

The constraints on new fermions follow from the EW measurements of the vector boson polarizations. The contributions of new quarks into polarizations of gauge bosons $\gamma,\,Z,\,W$
are described by Peskin-Takeuchi parameters (PT parameters). In our case, polarizations $\Pi_{ab}(0)=0$ and PT parameters can be represented as follows:
\begin{align}\label{2.6}
 S=&\frac{4s^2_w c^2_w}{\alpha}[\frac{\Pi_{ZZ}(M^2_Z,M^2_Q)}{M^2_Z}-\frac{c^2_w-s^2_w}{s_w c_w}\Pi^{'}_{\gamma Z}(0,M^2_Q)-\Pi^{'}_{\gamma\gamma}(0,M^2_Q)];
\,\,\,T=0;\notag\\
U=&-\frac{4s^2_w}{\alpha} [c^2_w\frac{\Pi_{ZZ}(M^2_Z,M^2_Q)}{M^2_Z}+2s_w c_w \Pi^{'}_{\gamma Z}(0,M^2_Q)+s^2_w\Pi^{'}_{\gamma\gamma}(0,M^2_Q)].
\end{align}
In Eqs.~(\ref{2.6}), $\alpha=e^2/4\pi$, $M_Q$ is mass of new quark and $\Pi_{ab}(p^2)$ are defined at $p^2=M^2_Z$ and $p^2=0$. The values $\Pi_{ab}(p^2,M^2_Q)$ can be described by the
expressions ($q=2/3$):
\begin{align}\label{2.7}
\Pi_{ab}(p^2,M^2_Q)=&\frac{g_1^2}{9\pi^2}k_{ab}F(p^2,M^2_Q);\,\,\,k_{ZZ}=s^2_w,\,k_{\gamma\gamma}=c^2_w,\,k_{\gamma Z}=-s_w c_w;\notag\\
F(p^2,M^2_Q)=&-\frac{1}{3}p^2+2M^2_Q+2A_0(M^2_Q)+(p^2+2M^2_Q)B_0(p^2,M^2_Q).
\end{align}
By straightforward calculations we get rather simple expressions for PT parameters:
\begin{equation}\label{2.8}
S=-U=\frac{k s^4_w}{9\pi}[-\frac{1}{3}+2(1+2\frac{M^2_Q}{M^2_Z})(1-\sqrt{\beta}\arctan\frac{1}{\sqrt{\beta}})].
\end{equation}
 In Eq.~(\ref{2.8}), $\beta=4M^2_Q/M^2_Z -1$, $k=16(4)$ (SQ model) with the value of charge $q=2/3(-1/3)$, and $k=20$ in the chiral-symmetric model. We check that the
 values of PT parameters significantly less the experimental limits \cite{PDG18}:
\begin{equation}\label{2.9}
S=0.00 +0.11(-0.10),\,\,\,U=0.08\pm 0.11,\,\,\,T=0.02+0.11(-0.12).
\end{equation}
 So, the scenarios with singlet and mirror quarks satisfy to the experimental EW restrictions on new physics.

 Additional EW restrictions follow from the flavor-changing neutral currents (FCNC). In the scenario, new quark does not mix with ordinary ones, FCNC are absent and there are no additional
 restrictions from the rare processes. Thus, the scenario with new heavy quarks is not excluded by precision EW restrictions. In Ref.~\cite{Bazhutov17},
it was shown that the potential of new meson and nucleon interaction has repulsive character. So, the DM particles do not form the coupled states with nucleons. This effect makes it
possible to escape strong cosmo-chemical constraints on the anomalous elements \cite{Bazhutov17}.

Quantum numbers, quark and isotopic structure of new hadrons are represented in Refs.~\cite{Bazhutov17,Beylin19}, where their properties and evolution are briefly described. Here, we
describe the principal properties of new mesons with structure of type $(qQ)$, in particular, the mesons $M=(M^0,M^-)$. The mass $M_0$ of neutral component $M^0$ is defined from the
equality of annihilation cross-sections at freez-out phase:
\begin{equation}\label{2.10}
(\sigma(M)v_r )^{Mod} =(\sigma v_r )^{Exp}
\end{equation}
In Eq.~(\ref{2.10}), the left part is model value of annihilation cross-section and the right part follows from the data on the recil abondence of DM, $(\sigma v_r)^{Exp}=2\cdot
10^{-9}\,GeV^{-2}$. The cross-section of annihilation $Q\bar{Q}\to gg,q\bar{q}$ was presented in \cite{Beylin19a}:
\begin{equation}\label{2.11}
(\sigma(M) )^{Mod}=\sigma(Q\bar{Q}\to gg,q\bar{q})\approx \frac{44\pi}{9}\frac{\alpha^2_s}{M^2}.
\end{equation}
Using the expression (\ref{2.11}) and equality (\ref{2.10}) we get the estimation of new quarks mass, $M\approx 10$ TeV. From this estimation, it follows that freezing out temperature
$T_f \approx M/30 \approx 300$ GeV, i.e., it is much greater than the temperature of QCD phase transition, $T_{QCD}\approx 150$ MeV. So, the stage of hadronization of ordinary and new
heavy hadrons begins much later the freezing out one. After phase transition new heavy quark $Q$ combine with ordinary light quark $q$ into new heavy $Q$-hadrons. In baryon
asymmetrical Universe it is possible the forming of meson states $(q\bar{Q})$ and baryon states $(qqQ)$ with unit electrical charge. Further, we consider the meson states only, while
the more complicated states were considered in Ref.~\cite{Beylin19a}.

\section{Interaction of DM with ordinary matter}

 Low-energy interaction of new hadrons with leptons is described by effective Lagranian in standard differential
form \cite{Kuksa20}:
\begin{equation}\label{3.1}
L^{eff}(WMM)=i G_{WM} U_{ik} W^{+\mu}(\bar{M}_{ui}\partial_{\mu} M_{dk} -\partial_{\mu}\bar{M}_{ui} M_{dk}) + h.c.,
\end{equation}
where $ui=u,c,t$; $dk=d,s,b$; $U_{ik}$ are the element of CM matrix, $M_{ui}=(ui\bar{U})$, $M_{dk}=(dk\bar{U})$, and effective coupling constant $G_{WM}=g/2\sqrt{2}$. The value of
$G_{WM}$ is equal to the coupling constant in $W$ -boson fundamental interaction with quarks. This is due to $W$-boson interacts with light standard quarks $u,d$ only, it does not
interact with heavy quark $Q$, which at low energy plays spectator role.

Low-energy Lagrangian of Z-boson interaction with new mesons can be represented in the form (\ref{3.1}) too. However, in contrast to $G_{WM}$, effective coupling $G_{ZM}$ is caused by
interactions of $Z$ with both quarks, $Q$ and $q$, and the problem of coupling definition arises.

Inelastic scattering of leptons on the $M$-particles is described by $t$-channel diagram with $W$-boson in the intermediate state. Using Eq.~(\ref{3.1}) and standard vertex $W
e^-\nu_e$, by straightforward calculation, in approximation $m_e \ll m(M)$ and $|m(M^-)-m(M^0)|\ll m(M)$, where $m(M)$ is the mass of new mesons, we get the cross-section in the form
\cite{Kuksa20}:
\begin{equation}\label{3.2}
\sigma(l^- M^0\to \nu_l M^-)\approx\frac{3g^4 |U_{ud}|^2}{2^{10}\pi M^4_W}s(1-\frac{\bar{m}^2}{s})^2,
\end{equation}
where $\sqrt{s}$ is full energy in the CMS and $\bar{m}$ is mean mass of the doublet ($M^0,M^-$). Full process of lepton scattering on $M^0$ with account of final states is:
\begin{equation}\label{3.3}
 l^- M^0\to \nu_l M^-\to \nu_l M^0 e^-\bar{\nu}_e.
\end{equation}
 So, in this process, neutrino with energy $E_{\nu}\sim E_l$ appears together with $e^-\bar{\nu}_e$ -pair. The cross-section of the process
 $\nu_l M^0 \to l^- M^+$ is described by the same expression (\ref{3.2}).

 Heavy DM particles are non-relativistic at the modern stage of evolution, they have an average velocity $\sim 10^{-3}$ with respect to Galaxy. From the kinematics of
the heavy-light particles collisions, when $m\ll M$, it follows that momentum transfer is small (see comments below). In this case, the low-energy DM-nucleon interaction can be
described by effective meson-exchange approach (see Ref.~\cite{Beylin20}). The nucleon-meson interaction was considered in \cite{Vereshkov91} on the base of the gauge scheme
realization of symmetry $U(1)\times SU(3)$. This scheme was developed and applied to the interaction of new heavy mesons with ordinary vector mesons \cite{Beylin20,Kuksa20}. Lagrangian
which describes the interaction of nucleons and new M-mesons with ordinary vector mesons consists of two terms:
\begin{equation}\label{3.4}
L_{NMV}=L_{NV} + L_{MV}.
\end{equation}
In Eq.~(\ref{3.4}) the first term describes interaction of nucleon with standard light mesons:
\begin{align}\label{3.5}
L_{NV}&=g_{\omega} \omega_{\mu}(\bar{p}\gamma^{\mu}p+\bar{n}\gamma^{\mu}n) + \frac{1}{2} g\rho^0_{\mu}(\bar{p}\gamma^{\mu}p-\bar{n}\gamma^{\mu}n)\notag\\
      &+\frac{1}{\sqrt{2}} g\rho^+_{\mu}\bar{p}\gamma^{\mu}n+\frac{1}{\sqrt{2}} g\rho^-_{\mu}\bar{n}\gamma^{\mu}p,
\end{align}
where $g_{\omega}=\sqrt{3}g/2\sin \theta$, $g^2/4\pi \approx 3.4$ and $\sin \theta\approx 0.78$ \cite{Vereshkov91}. The second term in Eq.~(\ref{3.5}) describes the interaction of M
-particles with ordinary vector mesons \cite{Kuksa20}:
\begin{align}\label{3.6}
L_{MV}&=iG_{\omega M}\omega^{\mu}(\bar{M}^0 M^0_{,\mu}-\bar{M}^0_{,\mu} M^0 +M^+_{,\mu}M^--M^+ M^-_{,\mu})\notag\\
      &+\frac{ig}{2}\rho^0_{\mu} (\bar{M}^0 M^0_{,\mu}-\bar{M}^0_{,\mu} M^0 +M^+_{,\mu}M^--M^+ M^-_{,\mu})\notag\\
      &+\frac{ig}{\sqrt{2}} \rho^{+\mu}(\bar{M}^0 M^-_{,\mu}-\bar{M}^0_{,\mu} M^-) + \frac{ig}{\sqrt{2}} \rho^{-\mu}(M^+ M^0_{,\mu}-M^+_{,\mu} M^0).
\end{align}
In Eq.~(\ref{3.6}), the coupling constant $G_{\omega M}=g_{\omega}/3$. In Ref.~\cite{Beylin20}, it was shown that scalar mesons give very small contribution into $NM$ interaction. The
interactions of new mesons with ordinary pseudoscalar mesons (for instance, $\pi$-mesons) are absent due to parity conservation. This is an important property which differs new heavy
hadrons from the standard baryons.

Low-energy scattering of nucleons on new mesons is described by $t$-channel diagrams with light vector and scalar mesons in the intermediate states. The diagrams with pseudoscalar
mesons are absent at the tree level, while the contribution of scalar mesons is negligible. So, the dominant contribution into the cross-section gives the change by the vector mesons,
$\omega$ and $\rho$ mesons.

Now, we consider the kinematics of elastic scattering $MN\to MN$, where $M=(M^0, M^-)$ and $N=(p,n)$. In the case of non-relativistc particles, the maximal value of momentum transfer
$Q^2=-q^2$ is $Q^2_{max}=(p k)^2\approx 4m^2_N v^2_r$. So, $Q_{max}\approx m_N v_r \sim 10^{-3}m_N$, the value $Q_{max}$ much less the mass of vector mesons $m_v$ ($m_v\sim m_N$) and
the meson-exchange model is relevant.

 Using the vertexes from the Eqs.~(\ref{3.5}) and (\ref{3.6}), we calculated the cross-section of the process $N_a M_b\to N_a M_b$ \cite{Beylin20,Kuksa20}:
\begin{equation}\label{3.7}
\sigma(N_a M_b\to N_a M_b) = \frac{g^4 m^2_p}{16\pi m^4_v}(1+\frac{k_{ab}}{\sin^2\theta})^2,
\end{equation}
where $N_a=(p,n)$, $M_b=(M^0,M^-)$, $g^2/4\pi\approx 3.4$, $\sin\theta=1/\sqrt{3}$ and $k_{ab}=\pm 1$ for the case of proton, $p$, and neutron, $n$. From the Eq.~(\ref{3.7}) it follows
that the value of cross-section is rather large, for example, $\sigma(pM^0\to pM^0)\approx 0.9$ barn. Large cross-section of $NM$-scattering can cause noteceabl interaction of DM halo
and galaxy. The problem of interaction between galaxies and their DM halo was considered in Ref.~\cite{Wechsler18}. Analysis of the low-energy scattering $N_a M_b \to N_a M_b$ discover
an important peculiarity of the $NM$-interaction. We show in Born approximation that potential of M-nucleon interaction at large distances ($d\sim m^{-1}_{\rho}$) has repulsive
character \cite{Bazhutov17,Beylin19a} and new heavy hadrons as DM-particles do not form coupled states with nucleon. This effect allows us to escape the problem of anomalous hydrogen
and helium \cite{Bazhutov17}.

In spite of large cross-section of $NM$-scattering, the direct detection of hadronic DM by underground devices is difficult due to small free pass in ground, $l_{fr}<1$ cm. So, we
consider indirect constraints on hadronic DM which can impact on the parameters of big bang nucleosynthesis (BBN) and $\gamma$-spectrum of cosmic rays (CR) in the Galaxy
\cite{Cyburt02}. In this work, the constraints were derived on the relation $R=\sigma(cm^2)/M_{DM}(g)$, where $\sigma$ is cross-section of DM-baryon scattering (in $cm^2$) and $M_{DM}$
is the mass of DM particle (in g). The constraints are as follow \cite{Cyburt02}:
\begin{equation}\label{3.8}
BBN:\,\,\,R < 10^8\, cm^2 g^{-1};\,\,\,\,\,\,CR:\,\,\, R<5\cdot 10^{-3}\,cm^2 g^{-1}.
\end{equation}
So, the second restriction is much more stringent and we compare it with model result. In our consideration, the value of mass is $M_{DM}\approx 10^4 \mbox{GeV}\approx 10^{-20}/0.56$
g, and cross-section $\sigma \sim 10^{-24} cm^2$. Thus, the model relation $R\approx 5.6\cdot 10^{-5}\,cm^2 g^{-1}$ does not contradict to the CR restriction. The more proper
measurements and constraints are considered in Ref.~\cite{Mack12} for the case of cosmic ray interaction with DM. The constrains were developed using NFW and Moore DM density profiles
and new data from Fermi gamma ray space telescope. Here, we use the upper constraint with Moore profile (which is more stringent): $\sigma_{Nx}=9.3\cdot 10^{-30} m_x cm^2 GeV^{-1}$. At
$m_x=10^4$ GeV we get $\sigma < 10^{-25} cm^2$, which excludes the model estimation. Here, we note that we describe the DM-nucleon interaction in meson-exchange approach using the
coupling constant, which was determined in low-energy hadrons interaction ($g^2/4\pi =3.4$). Thus, the model assumption concern the value of coupling is not justified, and from the
experiment we get the constraint on this parameter: $g^2/4\pi < 1$.

The processes of non-elastic scattering $N_a M_b \to N_c M_d$, where $N_a=(p,n)$ and $M_b=(M^0,M^-)$, are described by the kinematics of elastic scattering one. The dominant
contribution into cross-section is caused by $t$ -channel diagram with charged $\rho^{\pm}$-meson in the intermediate state. The expression for the cross-section explicitly indicates
the presence of the threshold:
\begin{equation}\label{3.9}
\sigma(N_a M_b\to N_c M_d)=\frac{g^4 m}{8\pi v_r m^4_v}\sqrt{2m}[E_a-\Delta_{ab}]^{1/2},
\end{equation}
where $E_a\approx m_a v^2_r/2$, $m(N_a)=m_a\approx m_b\approx m$, $\Delta_{ab}$ is the combination of mass-splitting $\Delta M=m(M^+) - m(M^0)$ and $\Delta m=m_n - m_p \approx 1.4$
MeV. The expression (\ref{3.9}) can be represented in another form:
\begin{equation}\label{3.10}
\sigma(N_a M_b\to N_c M_d)=\frac{g^4 m^2}{8\pi m^4_v}[1-\frac{\Delta_{ab}}{E_p}]^{1/2}.
\end{equation}
From (\ref{3.10}) one can see that the process of scattering has the threshold $E^{thr}_p = \Delta_{ab}$, when $\Delta_{ab}>0$. In Ref.~\cite{Kuksa20}, we present the expressions for
the threshold in the case of basic reactions, naimly $pM^0\to nM^+$, $nM^+\to pM^0$, $nM^0\to pM^-$ and $pM^-\to nM^0$.

\section{Fine and hyperfine splitting in doublet of new heavy mesons}

New heavy quarks possess strong QCD interaction, so they can form the coupled states - new heavy mesons, $(q\bar{Q})$, and fermions, $(qqQ)$, $(qQQ)$, $(QQQ)$. Classification and the
main properties of these hadrons were presented in \cite{Beylin19a} for the case of up and down type of new quark $Q$. The evolution of new hadrons was briefly considered in
Ref.~\cite{Bazhutov17}, where the process of burning out of heavy baryons was analyzed. Here, we represent the main properties of new mesons, $M^0$ and $M^-$, which can lead to the
characteristic sygnals of the hadronic dark matter.

An important role in hadronic DM scenario plays the value of mass-splitting in the doublet of neutral, $M^0$, and charged, $M^-$, new heavy mesons. We define the value of
mass-splitting as follows:
\begin{equation}\label{4.1}
\Delta m=m(M^-)-m(M^0).
\end{equation}
In the case of standard heavy-light (HL) mesons the value $\Delta m$ is an order of MeV, besides, this value is positive for the case of D-meson (up heavy quark) and negative for the
case of K- and B-mesons (down heavy quark). New heavy mesons $M^0$ and $M^-$ are just the case of the heavy-light (HL) mesons, $m_Q\gg m_q$. From the heavy quark symmetry
\cite{Isgur91}, it follows the analogy with standard HL mesons. So, for the case of up-type new mesons, $M^0=(u\bar{U})$ and $M^-=(d\bar{U})$, we can assume that $\Delta m$ is positive
and $\Delta m\sim$ MeV. The instability condition of charged meson $M^-$ leads to inequality $\Delta m>m_e$, where $m_e$ is the mass of electron. Thus, charged partner of neutral DM
particle has unique decay channel $M^-\to M^0W^{*-}\to M^0 e^-\bar{\nu}_e$ with very small phase space in a final state. The expression for the width of charged meson is as follows
\cite{Beylin19a}:
\begin{equation}\label{4.2}
\Gamma(M^-)=\frac{G^2_F}{60\pi^3}|M_{ud}|^2(\Delta m^5-m^5_e),
\end{equation}
where $M_{ud}$ is the element of CM matrix, which corresponds to the transition $d\to u W$. From the expression (\ref{4.2}) one can see that at $\Delta m\to m_e$, the value of width
$\Gamma(M^-) \to 0$, that is lifetime can be arbitrary large. For instance, at $\Delta m\sim 1$ MeV the lifetime $\tau \sim 10^5$ s. Thus, in the scenario with hadronic DM, new neutral
meson $M^0$, as DM candidate, has charged metastable partner with the same mass. New heavy charged meson appears in the process of collision of DM with ordinary matter, leptons and
nucleons (see the previous section).

Principal feature of hadronic DM scenario is the effect of hyperfine splitting of excited states of new heavy hadrons. In contrast to fine splitting, which is caused by change of light
quark content ($d\to u$) and has the value an order of MeV, hyperfine splitting takes place for the mesons with the same quark content and has much less value (an order of keV).
Further, we describe the effect of hyperfine splitting $\delta M_q=m(M^*_q)-m(M_q)$, where $M^*_q$ is excited state of the meson $M_q$. Here, we consider the lowest excited states of
the meson $M_q=(q\bar{U})$. In a direct analogy with the standard heavy-light (HL) mesons, $D_q=(cq)$ and $B_q=(\bar{b}q)$, we define the ground and excited states in the terms $S^1_0$
and $S^1_1$ (classification with quantum nubbers $L^{2s+1}_J$), or $\frac{1}{2}(0^-)$ and $\frac{1}{2}(1^+)$ (classification $I(J^P)$). Here, $L$, $s$, $J$, $I$ and $P=(-1)^{1+L}$ are
orbital momentum, spin, full momentum of the system, isospin and parity. The ground states $\frac{1}{2}(0^-)$ of the HL mesons we designate as $D_q$, $B_q$ and $M_q$, while the excited
states as $D^*_q$, $B^*_q$ and $M^*_q$. Evaluation of the mass-splitting of the states $M^*_q$ and $M_q$ we carry out in analogy with standard splitting mechanism. The analogy is
provided by the heavy quark symmetry which is the base of heavy quark effective theory (HQET). Heavy quark symmetry \cite{Isgur91} leads to relations between the masses of excited
states of $B$ and $D$ mesons \cite{Ebert98}:
\begin{equation}\label{4.4}
m(B_2)-m(B_1)\approx\frac{m_c}{m_b}(m(D_2)-m(D_1)),
\end{equation}
where $m(B_k)$ and $m(D_k)$ are masses of $B_k$ and $D_k$, $m_c$ and $m_b$ are masses of constituent quarks. The expression (\ref{4.4}) successfully describes the relation of splitting
between the lowest excited $\frac{1}{2}(1^-)$ and ground states $\frac{1}{2}(0^-)$ of $B$ and $D$ mesons:
\begin{equation}\label{4.5}
\frac{m(B^*)-m(B)}{m(D^*)-m(D)}\approx\frac{m_c}{m_b} \longrightarrow 0.32\approx 0.32\,(0.28).
\end{equation}
In (\ref{4.5}), we used $m(B^*)-m(B)=45$ MeV and $m(D^*)-m(D)=142$ MeV (see \cite{PDG18}), $m_c=1.55$ GeV and $m_b=4.88$ GeV \cite{Ebert98}. The value of relation in bracket (0.28)
follows from the data $m_c=1.32$ GeV and $M_b=4.74$ GeV \cite{PDG18}. In order to evaluate the mass-splitting in the doublet of new mesons $M_q=(q\bar{U})$, we use the relation
(\ref{4.5}) and equality $m(U)\approx m(M_q)=M$. Using the value of mass $M=10$ TeV, we get:
\begin{equation}\label{4.6}
\frac{\delta m(M)}{\delta m(B)}=\frac{m(M^*)-m(M)}{m(B^*)-m(B)}\approx\frac{m_b}{M} \longrightarrow \delta m(M)\approx\delta m(B)\frac{m_b}{M} \approx 2\, \mbox{KeV}.
\end{equation}
Thus, we get very small mass-spltiting (hyperfine splitting) $\delta m$, which is much less the fine splitting, $\delta m \ll \Delta m$.

The excitation of hadronic DM particle can manifest itself in the processes of interaction of neutral meson $M^0$ with cosmic rays. Transition to the first excited state of the meson
$M^0=(u\bar{U})$ can be caused by the absorption of photons in keV range, which have the wavelength $\lambda \sim 10^{-9}$ cm. If we assume that the meson $M^0=(u\bar{U})$ has the size
an order of nucleon radius, $R_M\sim 10^{-13}$ cm, then $R_M \ll \lambda_{trans}$ and interaction of $M^0$ with photons is caused by multi-pole expansion of charge distribution in the
system $(u\bar{U})$. So, the cross-section of $\gamma M^0$ scattering is small and these neutral mesons manifest themselves as dark matter particles. At $\lambda_{trans}\ll R_M$ the
cross-section of interaction $\gamma M^0$ become large and dark matter is not absolutely "dark". Now, we consider a possible manifestation of keV-signal, which is caused by hyperfine
splitting, in the spectrum of X-rays from the galaxy clusters. In Refs.~\cite{Bulbul14,Boyarsky14}, it was reported about emission line at $E\approx 3.5$ keV in a spectrum of galaxy
center and galaxy clusters. Here, we should note that the existence in nature of superheavy-light mesons inevitably (in the framework of HQET) leads to hyperfine mass-splitting of
ground and excited levels. Transitions between these states generate emission of photons with energy 3.5 keV when the mass of new heavy mesons $m(M)\approx 6$ TeV. This estimation in
the framework of HQET follows from the Eq.~(\ref{4.6}) without refer to DM hypothesis.

\section{Conclusion}

High-energy extensions of SM, as a rule, contain heavy particles which possess conservative quantum number. In the extension with singlet quark, the conservation of baryon charge leads
to the stability of the lightest new hadron which can be assumed as DM carrier. In this report, we present the main properties of new heavy hadrons and describe their low-energy
interactions with ordinary leptons and nucleons. We considered some electro-weak and cosmological constrains on the new heavy quarks and hadrons. Excited states of these hadrons were
considered in analogy with standard HL mesons. We show that there exist fine and hyperfine structure of excited levels which lead to weak or electromagnetic transitions. Such
transitions appear in the processes of interaction of new heavy hadrons with ordinary particles and cosmic rays. This effect rises the problem of separation the terms "dark matter" and
"hidden matter". It was noted that this problem becomes actual in the range of hard gamma rays.

In order to describe DM signals in hadronic processes, we developed and analyzed the low-energy model of new hadrons interaction with ordinary ones. The model of DM-nucleon interaction
is based on the meson-exchange approach and realized in the frame of the gauge scheme realization of the $SU(3)$-symmetry. In the framework of this model, we derived analytical
expressions for the cross-sections of elastic and inelastic collisions of nucleons and new heavy hadrons. These expressions will be used in the analysis and description of DM
interactions with cosmic rays, interstellar gas and with Earth atmosphere. The most important signal of such interactions is the appearance of the metastable heavy charge paticles
$M^-$. The scenario with hadronic DM provides a new aspect to the problem of interconnection of galaxies and their DM halos which can stipulate some pecuilarities of galaxy formation.
Effect of hyperfine splitting can explain the emission line at 3.5 keV in the spectrum of X-rays.

\section*{Acknowledgements}
Research was financially supported by Southern Federal University, 2020 (Ministry of Science and Higher Education of the Russian Federation) [In Gr /2020-03-IF].

%% The bibliography section

\end{document}